\documentclass{article}

\usepackage{amsmath,amssymb}
\usepackage{graphicx}
\usepackage{algorithm,algorithmic}
\usepackage{color}

\def\vect#1{\mbox{\boldmath{$#1$}}}
\def\mC{\mathbb{C}}

\def\mR{\mathbb{R}}

\newcommand{\Mm}{\vect M}

\def\Ac{{\cal A}}

\def\bfrho{\mbox{\boldmath$\rho$}}
\def\bfb{\mbox{\boldmath$b$}}
\def\dsp{\displaystyle}
\def\we{\vect e}

\newcommand{\bP}{{\bf {P}}}
\newcommand{\brho}{{\vect {\rho}}}
\newcommand{\bb}{{\vect {b}}}

\newcommand{\by}{\vec{\bf y}}

\def\norm#1{| #1 |}

\newcommand{\sign}{\text{sign}}

\newtheorem{remark}{Remark}

\begin{document}

\title{Synthetic aperture imaging with intensity-only data} 


\author{Miguel Moscoso%
\thanks{Department of Mathematics, Universidad Carlos III de Madrid, Leganes, Madrid 28911, Spain ({moscoso@math.uc3m.es})}, 
\and
Alexei Novikov%
\thanks{Mathematics Department, Penn State University, University Park, PA 16802 ({novikov@psu.edu})},
\and
George Papanicolaou%
\thanks{Department of Mathematics, Stanford University, Stanford, CA 94305 ({papanicolaou@stanford.edu})},
\and
Chrysoula Tsogka%
  \thanks{Applied Math Unit, University of California, Merced, 5200 North Lake Road, Merced, CA 95343 (l{ctsogka@ucmerced.edu}).}%
 }

\maketitle

\begin{abstract}
We consider imaging the reflectivity of scatterers from intensity-only data recorded by a single moving transducer that both emits and receives signals,
forming a synthetic aperture. By exploiting frequency illumination diversity, we obtain multiple intensity measurements at each location, from which we 
determine  field cross-correlations using an appropriate phase controlled illumination strategy and the inner product polarization identity. The field
cross-correlations obtained this way do not, however, provide all the missing phase information because they are determined up to a phase that depends on the receiver's location. 
The main result of this paper is an algorithm with which 
we recover the field cross-correlations up to a single phase that is common to all the data measured over the synthetic aperture, so all the data are synchronized. 
Thus, we can image coherently with data over all frequencies and measurement locations as if full phase information was recorded.
\end{abstract}

\maketitle

\section{Introduction} 
{
We consider a multifrequency phaseless synthetic aperture imaging system composed of a single transmitter/receiver element that operates at microwave frequencies. 
The system only records the intensities of the signals, and forms the images by combining the data coherently over the entire synthetic aperture. 
With the proposed computational imaging approach, we show that using intensity-only data is as good as coherent imaging with full (phase and amplitude) data; cross-range and range resolutions are still
inversely proportional to the synthetic aperture size and to the bandwidth of the recorded signals, respectively.}

{
Imaging with microwaves is of interest in security applications because they can penetrate through materials that are opaque at visible wavelengths, allowing the detection of concealed objects under cloths and inside luggage \cite{Martinez12, Sheen01}. Microwaves have also been successfully used  for through-wall imaging \cite{Wang12},  breast cancer detection in medical imaging  \cite{Nikolova11, Kwon16}, and space surveillance \cite{Czerwinski14}. An arbitrary number of transmitters and receivers is used in general, typically arranged in a planar or spherical geometry, while a moving single transmitter/receiver element can also be used so as to form a synthetic aperture. This last mode records less information but it can still provide data with enough frequency and space diversity for imaging.}

{
In all applications, control and detection of phases is essential to image coherently for good resolution.  However, to maintain phase coherence during the whole data acquisition process may be difficult or impossible.
This is the case, for example, when high frequency signals, above  $30$ GHz, and large synthetic apertures are used to form high quality images \cite{Laviada15,Yurduseven17}. 
Other situations in which phase measurements may not be reliable arise when there is uncertainty in the antenna location or in the signal trajectory.}

{The conventional way to overcome lack of coherent phase measurements is to resort to computational imaging
systems that use intensity-only data recorded with simpler system architectures. The missing phases are then reconstructed
with the well-known phase retrieval algorithms, which rely in an essential way on prior information about the object to be imaged.
This is the case in  various  fields including  crystallography,  optical imaging,  astronomy,  and  electron  microscopy.
}

{
We recount briefly the basic facts about phase retrieval. {In this problem, the objective is to reconstruct a signal from its power spectrum. This is, of course, an ill-posed
problem because there is not enough information to recover the true signal. To resolve this issue, one can invoke prior knowledge about the signal and look for a solution by using a phase retrieval algorithm.} The most widely used algorithms 
are the alternating projection algorithms introduced by Gerchberg and Saxton \cite{GS72} and by Fienup \cite{Fienup82, Fienup13}
that project the iterates on intensity data (or prior information) sequentially in both the real and the Fourier spaces. Although these algorithms are efficient and flexible for reconstructing
the missing phases in the data, and performance is often good in practice, they may not converge or even get close to the true, missing phases because the problem is non-convex.  

{In \cite{Soldovieri05}, the authors use the quadratic approach which formulates the phase retrieval problem as a non-linear (quadratic) inverse problem with data the square amplitude of the near field. For this quadratic inverse problem, the presence of local minima can be avoided by increasing the number of independent data (cf. \cite{Soldovieri05} and references therein)}. 
 A convex, non-iterative approach that guarantees exact recovery {in the case of sparse reflectivity} was proposed in \cite{CMP11,Candes13}, but its computational cost is  high when the problem is large.
}

{
Holographic methods can also be used. These are interferometric approaches that obtain phase field differences from  intensity data.  For example, phase information can be recovered by superposing the scattered signals with a known and well-controlled  reference signal. 
 Holography was first introduced 
 to increase the resolution of electron microscopes \cite{Gabor49}. It was a two-step method for recording the phase of optical signals, since photographic film is not sensitive to complex amplitudes but to intensities. Later, holography was used with microwaves for measuring antenna radiation patterns \cite{Leith62,Bennet76}. For recent adaptations of microwave holographic techniques, we refer to \cite{Costanzo05,Laviada13,Smidth14,Laviada15}.}
 
{
On the other hand, the Wirtinger Flow (WF) phase retrieval algorithm of \cite{CandesLiSolt15} is used  in \cite{Fromenteze16} for microwave imaging with a frequency-diverse metasurface antenna. The  antenna produces spatially diverse radiation patterns that vary as a function of the frequency sampled over the operational K-band (17.5-26.5 GHz). In \cite{Yurduseven17}, the authors use the more recent sparse WF algorithm  proposed in \cite{Yuan17} that allows to reduce the computational cost of the method. } 

{
We follow here a different strategy inspired by interferometry.  The key idea is {to resolve the non-uniqueness of the phase retrieval problem creating redundancy in the data by illuminating
the image multiple times. 
Indeed, by using an appropriate illumination strategy and the inner product polarization identity, the missing phase information can be uniquely determined, up to a global phase~\cite{Novikov14,Moscoso16,Moscoso17}. The polarization identity is a well known formula in mathematics that relates the inner product of two vectors with their norms. It is not related to the polarization of electromagnetic waves.}
Special forms of the polarization identity have also been used in \cite{Lehto92} where antenna phase patterns are obtained from the responses to two probe antennas recorded at three power detectors, and in  \cite{Costanzo05}  where intensity data are collected with two probes at fixed offset moving over an arbitrary scanning surface.}

In this paper, we present a new computational imaging approach to accurately reconstruct the reflectivity of scatterers with synthetic aperture, intensity-only data.  
The method has two stages. First,  from the intensity data at each source-detector
position, we recover field cross-correlations corresponding to coherent
sources of different frequencies. This is achieved using a special sequence of illuminations that exploits the frequency diversity available on the transmitter side. 
The recovered field cross-correlations are the same as the ones obtained from full data, up to a phase that is different at each receiver location. Hence, 
at this point, these field cross-correlations cannot be combined coherently to image the reflectivity. {To use them coherently over all the
synthetic aperture they need to be synchronized or aligned first.}

{This is the second step of the proposed method in which all the phases that depend on
the receiver locations are referred to a single global phase. 
The main idea of the second step is to refer the total reflectivity estimated at each receiver location to the total true reflectivity, which is 
a common quantity for all measurement locations. This is the main contribution of the paper. 
With the strategy proposed here, we show that imaging with intensity-only data is as good as imaging with full-phase data.}

The paper is organized as follows. In Section \ref{sec:holo}, we explain the proposed method to obtain coherent cross-correlations when 
 one element that transmits multifrequency microwaves and measures only intensities is used to collect the data on a synthetic aperture. 
 In Section \ref{sec:methods}, we discuss
 two imaging methods traditionally used when full (phase and amplitude) data are available. We use these methods with the recovered cross-correlations.
 In Section \ref{sec:results}, we show the results of our numerical experiments. Section \ref{sec:conclusions} contains our conclusions.

\section{Multi-frequency interferometric synthetic aperture imaging} \label{sec:holo}
\begin{figure}[htbp]
\begin{center}
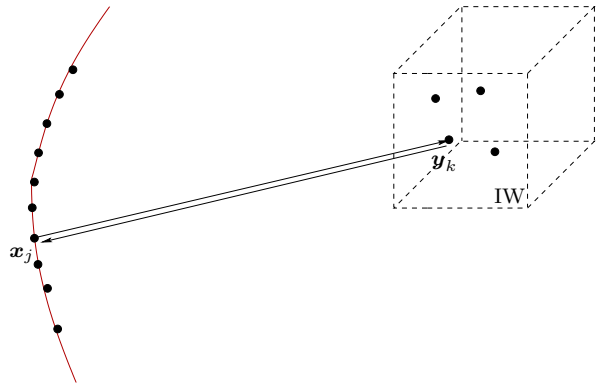
\caption{General setup of a synthetic aperture imaging problem. The transducer at $\vect x_j$  emits a probing signal and the reflected intensities are recorded at the same location for all illuminations. The scatterers are located inside the image window IW which is discretized with the grid points ${\vect y}_k$, $k=1,\ldots,K$.}
\label{fig:setup}
\end{center}
\end{figure}

{In the next two subsections we present in detail the two stages of the proposed approach.
The goal is to determine the reflectivity $\bfrho$ within a region of interest, called the image window IW, from multiple  intensity-only measurements at different  locations $\vect x_j$, $j = 1, \ldots, N$, and frequencies $\omega_l$, $l=1,\ldots, S$, with a total number of $D=N\cdot S$ data (see Figure \ref{fig:setup}). We denote by $|P(\vect x_j,\omega_l) |$  the amplitude of the signal received at location $\vect x_j$ when a signal of frequency $\omega_l$, unit amplitude, and zero phase is emitted from the same location.}

{For imaging purposes, the IW is discretized using a uniform grid of $K$ points ${\vect y}_k$, $k=1,\ldots,K$.  The unknown is the {\it reflectivity vector} $\vect\rho=[\rho_{1},\ldots,\rho_{K}]^t\in\mC^K$, whose entries are the values of the reflectivity $\rho_k=\rho({\vect y}_k)$ on the grid points ${\vect y}_k$, $k=1,\ldots,K$. We assume that $K>D$, and often we have $K\gg D$. Moreover, we assume that the unknown reflectivity vector is $M$-sparse with $M \ll K$. This is often true in applications where the reflectivity to be imaged does not occupy the entire scene but rather a small part of the IW.}

\subsection{Multi-frequency field cross-correlations}
We pursue here the idea of  \cite{Novikov14, Moscoso16, Moscoso17}, where it is shown that {field} cross-correlations can be obtained from
intensity-only measurements by using an appropriate protocol of illuminations and the polarization identity. {In these works, 
the illumination strategy was implemented with an array in which all its elements were used  to emit  and receive signals. 

In synthetic aperture imaging, however,  there is an inherent loss of possible illuminations because only one 
transmitter/receiver element is used. This lack of flexibility, or diversity, in illuminations might at first suggest that the  data cannot 
be used coherently to form images 
when only intensities are recorded.}

To consider this issue further, we introduce the  row vector 
\begin{equation}
 \bP^j =\left[ P(\vect x_j,\omega_1) \quad
  P(\vect x_j,\omega_2) \quad \ldots \quad
  P(\vect x_j,\omega_{S})
  \right]  
\end{equation}
with $S$ components. The entry $P(\vect x_j,\omega_l)$ corresponds to the signal recorded  at $\vect x_j$, including phases, when a unit amplitude and zero phase signal  of frequency
$\omega_l$ is sent from the same location $\vect x_j$.
{The full phase retrieval problem consists on determining, at once, the $N \times S$ phases of the phaseless measurements   $\left| P(\vect x_j,\omega_l) \right|$ for the $N$ measurement positions and the $S$  frequencies.}

{In this paper, we use a different strategy. First, we recover  the field cross-correlation matrices at each receiver, up to a phase that depends on the receiver location but
not on the frequency. In a second step, all these phases are referred to a single one; and this allows for coherent imaging.}

Specifically, the first step of the proposed methodology consists on recovering, at every fixed location $\vect x_j$, the {field} cross-correlation matrix 
\begin{eqnarray}
\label{CC1}
  [\Mm^j]_{ll'}\equiv\dsp m^j_{ll'} = \overline{P(\vect x_j,\omega_l)}
  P(\vect x_j,\omega_{l'}) , 
\end{eqnarray}
for $l,l'=1,\dots,S$.
To do so, we use the diversity over illuminations with different frequencies, and the {inner product} polarization identity
\begin{eqnarray}\label{polarization_re}
\mbox{Re}(m^j_{ll'})= 
\frac{1}{2} \left( \norm{\bP^j \cdot \we_{l+l'}}^2  - 
\norm{\bP^j \cdot \we_l}^2 - \norm{\bP^j \cdot \we_{l'}}^2\right) \, ,
\end{eqnarray}
\begin{eqnarray}\label{polarization_im}
\mbox{Im}(m^j_{ll'})= 
 \frac{1}{2} \left( \norm{\bP^j \cdot \we_{l-{\imath} l'}}^2  -
  \norm{\bP^j \cdot  \we_l}^2 - \norm{\bP^j \cdot \we_{l'}}^2\right)\, .
  \end{eqnarray}
Here, $\we_{l} \in \mathbb{C}^S$ is the vector with  $1$ in the $l$-th coordinate and $0$'s elsewhere. It represents a signal of unit amplitude and zero phase at frequency $\omega_l$. 
In \eqref{polarization_re}-\eqref{polarization_im}, 
${\imath}= \sqrt{-1}$, $\we_{l+l'}= \we_{l}+\we_{l'}$,  and $\we_{l-{\imath} l'}= \we_{l}-{\imath} \we_{l'}$. 
The vector 
$\we_{l+l'}=\we_l+\we_{l'}$ refers to sending simultaneously signals of unit amplitude and zero phase at two frequencies $\omega_l$ and $\omega_{l'}$, 
while the vector $-{\imath} \we_{l'}$ denotes a $-\pi/2$ phase shift in the signal of frequency $\omega_{l'}$.

When two signals of  distinct frequencies $\omega_l$ and $\omega_{l'}$ are sent simultaneously to probe the medium, the intensity at the receivers oscillates at the difference frequency 
$\omega_l - \omega_{l'}$. Therefore, our imaging system should be able to resolve intensities that oscillate at frequencies of the order of the total bandwidth $B$, which determines the range resolution $c/B$ of the imaging system, where $c$ is the signal speed. {In Appendix \ref{app:meas}, we describe in detail the measurement process with which we  recover all the elements of the cross-correlation matrix $\Mm^j$.}
{We do not assume, however,  cross-correlation measurements between different positions $\vect x_j$ and, hence,
we recover $\Mm^j$ up to an unknown global phase $\theta_j$, {\em i.e.,} a factor of the form $e^{i \theta_j}$, with $\theta_j$ depending on  $\vect x_j$. 
}

{This means that, in principle,
we can only recover asynchronized signals between the measurement locations. 
We remark that synchronization is needed regardless of whether the elements in $\Mm^j$ are 
obtained through \eqref{polarization_re}-\eqref{polarization_im}. It is a problem that arises due to the lack of information between signals sent and received at different positions. 
Hence, the cross-correlations over frequencies need to be synchronized or aligned over the receiver positions to image coherently.}

{
This is the main difficulty in synthetic aperture imaging when the phases are not measured. We explain in the next subsection how to synchronize the signals in the frequency domain, i.e., 
how to recover $N$ individual phases  from  phase difference measurements of the form \eqref{CC1}.
Once all the signals are synchronized, the imaging problem is trivial as we show in Section \ref{sec:methods}.
} 
 
 We describe in Appendix \ref{app:illum} an illumination strategy that requires 
 $3S-2$ illuminations to recover all the entries in $\Mm^j$.  {This is the minimum number of measurements needed per receiver location.}

\subsection{Location-dependent phase recovery}  \label{sec:phase}
As discussed above,  we can recover, up to a global phase that depends on the receiver location $\vect x_j$,  {field} cross-correlations  of the form \eqref{CC1} 
using the frequency diversity in the illuminations. Since the amplitudes $|P(\vect x_j,\omega_{l})|$ are  known at every location $\vect x_j$ for all the 
frequencies $\omega_l$, we can compute

$$\frac{m^j_{ll'}}{ | P(\vect x_j,\omega_l)|} = P(\vect x_j,\omega_{l'}) \frac{\overline{P(\vect x_j,\omega_l)}}{ | P(\vect x_j,\omega_l)|},\ l'=1,\ldots,S. $$
This means that full data can be recovered at each measurement  location $\vect x_j$ up to a global phase $\theta_j$ which is unknown. In other words,
\begin{equation}
  b^j_l= P(\vect x_j,\omega_{l}) e^{\imath \theta_j}, \mbox{ for } l=1,\ldots,S
\label{eq:dataj}
\end{equation}
is known to us.
To  image coherently, we have to refer all the unknown phases $\theta_j$  to a single location. {This is a 
{synchronization problem} with $N$ unknown phases instead of $N\times S$ unknown phases as for the full phase retrieval problem. We next explain how to solve this problem.} 

{
\begin{remark}We assume here that the measurements $|P(\vect x_j,\omega_l)|$ are kept only when they are above some threshold determined by 
the noise level. Otherwise, they are not used. In the numerical simulations discussed below it was not necessary to discard any measurements, even in the $0$dB case.\end{remark}}

\begin{figure}[htbp]
\begin{center}
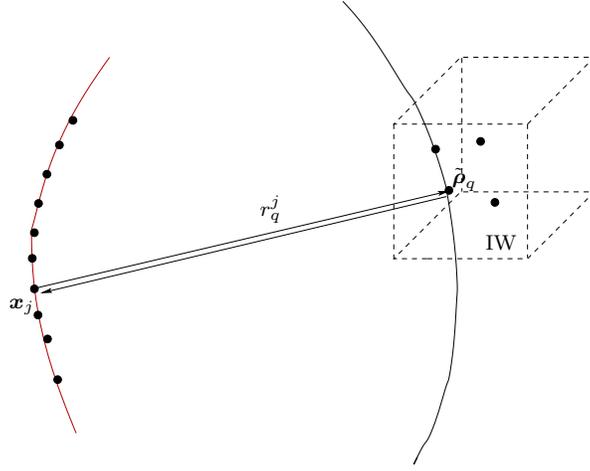
\caption{The data (with phases) $P(\vect x_j,\omega_{l})$ recorded at transducer $\vect x_j$  when a unit amplitude signal with zero phase at frequency $\omega_l$ is emitted from the same location is given by the model \eqref{model1d} where  $\tilde{\rho}_q$ is the integral of the unknown reflectivity on the sphere centered at $\vect x_j$ of radius $r^j_q$. }
\label{fig:setup2}
\end{center}
\end{figure}

By linearizing  the scattering problem and assuming that multiple scattering is negligible,  the data is given by
\begin{equation}
P(\vect x_j,\omega_{l}) = \sum_{q=1}^{Q} \tilde{\rho}_q e^{\imath 2 \frac{\omega_l}{c} r^j_q},
\label{model1d}
\end{equation}
where $c$ is the velocity in a homogeneous medium, and $\tilde{\rho}_q$ is the integral of the reflectivity on the sphere of radius $r^j_q$ centered at $\vect x_j$; see Figure \ref{fig:setup2}. Typically for sparse $\bfrho$,  $\tilde{\rho}_q$ is a reflectivity of a single scatterer.  {We assume here that the reflectivity is frequency independent.} 
It follows from \eqref{model1d} that the data  $P(\vect x_j,\omega_{l})$ is the Fourier coefficient of the reflectivity $\tilde{\brho}$ corresponding to wavenumber $\kappa_l= 2 \frac{\omega_l}{c}$, i.e., 
 $$ \widehat{{\rho}_j}(\kappa_l) = \sum_{q=1}^Q \tilde{\rho}_q e^{\imath \kappa_l r^j_q}.$$
Therefore, at each source-receiver position we have to solve a one dimensional problem to recover $\tilde{\rho}_q$, $q=1,\ldots,Q$, from the processed data \eqref{eq:dataj}. Again, these are data with phases that are well defined up to global phases $\theta_j$ that are not known. {This means that the data are trains of asynchronous spikes, each train corresponding to a measurement location $\vect x_j$.}
 
Still, we can determine the vector $\tilde{\bfrho}=[\tilde{\rho}_1,\tilde{\rho}_2,\dots,\tilde{\rho}_Q]$ by solving the following linear system 
\begin{equation}
 \Ac^j \tilde {\brho}^j  = \bb^j,
\label{s1d}
\end{equation}
with sensing  matrix 
\begin{equation}
\Ac^j = \left[
\begin{array}{cccc}
e^{\imath 2\frac{\omega_1}{c} r^j_1} & e^{\imath 2 \frac{\omega_1}{c} r^j_2} & \cdots & e^{\imath 2 \frac{\omega_1}{c} r^j_Q} \\
e^{\imath 2 \frac{\omega_2}{c} r^j_1} & e^{\imath 2 \frac{\omega_2}{c} r^j_2} & \cdots & e^{\imath 2 \frac{\omega_2}{c} r^j_Q} \\ 
\vdots & & & \vdots \\
e^{\imath 2\frac{\omega_S}{c} r^j_1} & e^{\imath 2 \frac{\omega_S}{c} r^j_2} & \cdots & e^{\imath 2 \frac{\omega_S}{c} r^j_Q} \\
\end{array}
\right],
\label{Acj}
\end{equation}
and data vector $\bb^j$ with components given by \eqref{eq:dataj}.
The superscript $j$ is used to stress that the linear systems \eqref{s1d} uses data recovered at location $\vect x_j$.

{These linear systems are underdetermined and, hence, there are infinitely many scatterer's configurations that match the data. However, if  the true reflectivity $\bfrho_0$  is sparse, with only a few components different than zero, an
$\ell_1$-minimization approach can find their unique sparse solution.
Exact recovery is guaranteed under  the assumption that the mutual coherence\footnote{The mutual coherence of $\Ac$ is defined as $\max_{i \ne j}  
|\langle \vect a_i,\vect a_j \rangle |$ with $\vect a_i \in \mC^N$ the columns of $\Ac$ normalized to one, so that $\|\vect a_i \|_{\ell_2} =1$ $\forall \, i=1,\ldots,K$.}   of the matrices $\Ac^j$ 
are smaller than $1/(2M)$, where $M$ is the number of non zero components in a vector $\tilde {\brho}^j$. For more details about  $\ell_1$-minimization methods we refer to \cite{Donoho03, Candes06a, Fannjiang12b, Borcea15}. 
In the simulations shown below, we use a generalized lagrangian multiplier algorithm (GeLMA)~\cite{Moscoso12}, described in Algortihm \ref{algo}, to find the sparsest solution to \eqref{s1d}. 

The matrix $\Ac^j$ defined in \eqref{Acj} depends on the radii $r^j_i$, $i=1,\ldots,Q$, which are computed in the following way. Given an IW with discretization points $\vect y_k$, $k=1,\ldots,K$, 
we compute the distances from all points $\vect y_k$ to the receiver location $\vect x_j$, 
\begin{equation} \label{dist}
 {R}_k^j = | \vect y_k - \vect x_j | .
 \end{equation}
These form the components of a vector in $\mR^K$. We then sort the components of this vector in ascending order and keep only the entries that appear with multiplicity larger than one. In practice, we only keep the entries that differ from each other by at least a level $\epsilon$. The value of $\epsilon$ should be small enough so  we do not disregard many components since that would affect the accuracy of the reconstruction, but it cannot be very close to zero because we do not want the columns of $\Ac^j$ to be almost parallel.  {Note that $\epsilon$ has units of length and should be chosen to be small with respect to the wavelength and the pixel size, so that neglecting distances that differ less than $\epsilon$ does not affect the accuracy of the recovered phases.} Assuming the $R_k^j$ are ordered then this can be done as follows, 
\begin{equation}
\label{Rjk}
\begin{array}{l}
\dsp \mbox{set } i=1 \mbox{ and } r_i^j=R_i^j \\
\mbox{for k=2 to K} \\
\hspace*{0.5cm} \mbox{if }|R_k^j-R_{k-1}^j| > \epsilon \\
\hspace*{0.7cm} \mbox{set } i=i+1 \mbox{ and }  r_i^j=R_k^j \\
\hspace*{0.5cm} \mbox{end}\\
\mbox{end}
\end{array} 
\end{equation}
 This process generates the radii $r_i^j$, $i=1,\ldots,Q$, with $Q\le K$, that depend on the receiver locations $\vect x_j$, $j=1,\ldots,N$.

Once the solution vectors $\tilde {\brho}^j$ have been found, we compute the total reflectivity within the IW 
by summing all the components of the vectors $\tilde{\brho}^j$. That is, for each receiver location we compute the scalar
\begin{equation}
 \sum_{q=1}^Q \tilde{\rho}^j_q \approx e^{\imath \theta_j} \frac{1}{h_c} \int_{IW} \brho_0 d\by,
\label{eq:Arhoj}
\end{equation}
with $h_c$ a constant that depends on the discretization.  
The key point here is that for all receiver positions we can compute an approximation to
the total reflectivity {$\int_{IW} \brho_0 d\by$},  up to unknown phase factors $ e^{\imath \theta_j}$, $j=1,\ldots,N$. 

Thus, we can refer all the recovered quantities \eqref{eq:dataj} to a 
same global phase with no physical meaning for imaging purposes. Indeed, let us define the quantities
\begin{equation}
c_j = \frac{ \sum_{q=1}^Q  \tilde{\rho}^j_q} {\sum_{q=1}^Q  \tilde{\rho}^1_q} \stackrel{\eqref{eq:Arhoj}}{=} e^{\imath (\theta_j -\theta_1)} , \ j=1,\ldots, N,
\label{eq:cj}
\end{equation}  
by dividing the total reflectivities associated to every location $\vect x_j$  by the total 
reflectivity obtained from the measurements recorded at $\vect x_1$.  The choice of $j=1$ in the denominator in 
\eqref{eq:cj} is, of course,  arbitrary. With this choice, $c_1=1$.
Then, by multiplying the recovered data \eqref{eq:dataj} by $\overline{c_j}$ we get
\begin{equation}
 \overline{c_j} b^j_l = P(\vect x_j,\omega_{l}) e^{\imath \theta_1}, \,\, \forall \, j=2, \ldots, N \, \mbox{and}  \,\, \forall \,\, l=1,\ldots,S.
\label{eq:datajcorrect}
\end{equation}
This defines the holographic data
\begin{equation}
\begin{array}{ll}
\dsp P^h(\vect x_1,\omega_{l}) &\dsp = b^1_l,  \,\,  \forall  \,\,  l=1,\ldots,S.\\
\dsp P^h(\vect x_j,\omega_{l}) &\dsp =  \overline{c_j} b^j_l,  \,\, \forall \, j=2, \ldots, N \, \mbox{and}\,\, l=1,\ldots,S.
\end{array}
\label{eq:datah}
\end{equation}
The phases in \eqref{eq:datah} are now coherent over different receiver positions and frequencies! Thus, the unknown reflectivity $\brho$ can be reconstructed as if data with phases were recorded. 

We want to emphasize that the proposed methodology allows one to produce holographic data from intensity measurements. This is of considerable importance since: i) intensity data are much easier to obtain, and can be recorded with less expensive equipment (sensors) than data obtained with holographic techniques, ii) holographic data contain coherent phase information and allow us to obtain depth resolved reconstructions,  and  iii) the proposed methodology does not need any prior information about the sought reflectivity.    
We compare next the performance of different imaging methods using \eqref{eq:datah} as  data. 
\section{Full phase synthetic aperture imaging methods}
\label{sec:methods}
Once the holographic data \eqref{eq:datah} are obtained, the unknown reflectivity can be reconstructed with any imaging method as if the data with phases were recorded with a synthetic array aperture. Here we show results obtained  with the frequently used Kirchhoff migration (KM) imaging method and the $\ell_1$-optimization approach. 

KM is a direct imaging method which can be written as
\begin{equation}
{\rho}^{KM}({\vect y}_k)=\sum_{j=1}^N \sum_{l=1}^S
e^{- \imath 2 \frac{\omega_l}{c} | \vect x_j - \vect y_k |}  P^h(\vect x_j, \omega_l ),
\label{eq:KM_sar}
\end{equation}
where $| \vect x_j - \vect y_k |$ is the distance between the measurement location $\vect x_j$ and the search point $\vect y_k$ in the IW. The image ${\bfrho}^{KM}=[{\rho}^{KM}({\vect y}_1),{\rho}^{KM}({\vect y}_2)\dots,{\rho}^{KM}({\vect y}_K)]$ is an approximation to the true reflectivity vector $\bfrho_0$. 

We also form an image by promoting  a sparse solution to the  linear system 
\begin{equation}\label{system}
\Ac \bfrho=\bfb\, ,
\end{equation}
where  $\bfrho \in \mC^{K}$ is the sought reflectivity vector,  
\begin{equation}
\Ac = \left[
\begin{array}{cccc}
e^{\imath 2\frac{\omega_1}{c} | \vect x_1 - \vect y_1 |} & e^{\imath 2 \frac{\omega_1}{c} | \vect x_1 - \vect y_2 |} & \cdots & e^{\imath 2 \frac{\omega_1}{c} | \vect x_1 - \vect y_K |} \\
e^{\imath 2\frac{\omega_1}{c} | \vect x_2 - \vect y_1 |} & e^{\imath 2 \frac{\omega_1}{c} | \vect x_2 - \vect y_2 |} & \cdots & e^{\imath 2 \frac{\omega_1}{c} | \vect x_2 - \vect y_K |} \\
\vdots & \vdots & & \vdots \\
e^{\imath 2\frac{\omega_1}{c} | \vect x_N - \vect y_1 |} & e^{\imath 2 \frac{\omega_1}{c} | \vect x_N - \vect y_2 |} & \cdots & e^{\imath 2 \frac{\omega_1}{c} | \vect x_N - \vect y_K |} \\
e^{\imath 2\frac{\omega_2}{c} | \vect x_1 - \vect y_1 |} & e^{\imath 2 \frac{\omega_2}{c} | \vect x_1 - \vect y_2 |} & \cdots & e^{\imath 2 \frac{\omega_2}{c} | \vect x_1 - \vect y_K |} \\
\vdots & \vdots  & & \vdots \\
e^{\imath 2\frac{\omega_S}{c} | \vect x_N - \vect y_1 |} & e^{\imath 2 \frac{\omega_S}{c} | \vect x_N - \vect y_2 |} & \cdots & e^{\imath 2 \frac{\omega_S}{c} | \vect x_N - \vect y_K |} \\
\end{array}
\right]
\label{Ac}
\end{equation}
is the model matrix, and $\bfb \in \mC^{N \cdot S}$  is the recovered data vector whose components are
\begin{equation}
b_i=b_{(l-1)N+j} = P^h(\vect x_j, \omega_l ),\,\, j=1,\ldots,N, l=1,\ldots,S.
\label{eq:defb}
\end{equation}
We note that the KM solution \eqref{eq:KM_sar} can also be written as  $\bfrho^{KM} = \Ac^* \bfb$,
where $\Ac^*$ is the complex conjugate transpose of $\Ac$. 

To find the sparsest solution to the system \eqref{system}, we solve the $\ell_1$-minimization problem 
\begin{equation}\label{eq:ell1}
\min\|\bfrho\|_{\ell_1}\quad\text{subject to }\quad \Ac\bfrho=\vect b\, ,
\end{equation}
using GeLMA described in Algorithm \ref{algo}. This  algorithm involves  matrix-vector multiplications followed by a shrinkage-thresholding step defined by the operator 
$$\eta_\tau(y_i)=\sign(y_i)\max\{0,|y_i|-\tau\}.$$ 
{GeLMA converges to the solution of (\ref{eq:ell1}) independently of the regularization parameter $\tau$, see \cite{Moscoso12}.}

\begin{algorithm}
\begin{algorithmic}
\REQUIRE Set $\vect y =\vect0$, $\vect z=\vect 0$. Pick the step size $\beta$, and a regularization parameter $\tau$.
\REPEAT
\STATE Compute the residual $\vect r= \vect b- \Ac\vect y$ 
\STATE $\vect y \Leftarrow  \eta_{\tau\beta}(\vect y + \beta\Ac^\ast(\vect z+ \vect r))$
\STATE $\vect z \Leftarrow\vect z + \beta\vect r$
\UNTIL{Convergence}
\end{algorithmic}
\caption{GeLMA for solving \eqref{eq:ell1}}
\label{algo}
\end{algorithm}
As discussed previously, the solution of (\ref{eq:ell1}) agrees with the exact $M$-sparse solution $\bfrho$ for noiseless data if the mutual coherence of the matrix $\Ac$ is smaller than $1/(2M)$.

\section{Numerical simulations}
\label{sec:results}
We consider a high frequency microwave scanning regime with central frequency $f_0=50$GHz which corresponds to $ \lambda_0=6$mm. We make measurements for $S=41$ equispaced frequencies covering a total bandwidth of $10$GHz using a single transmitter/receiver that is moving along a linear trajectory. The synthetic aperture is $a=20$cm, and the distance from its center to the center of the IW is $L=1$m; see Figure \ref{fig:geometry}. 
We assume that the medium between the synthetic array and the IW is homogeneous. The size of the IW is $48\mbox{cm} \times 48\mbox{cm}$, and the pixel size is $6\mbox{mm}\times  6\mbox{mm}$. The measurements are gathered at $N=41$ equispaced locations.  These parameters are typical in  microwave scanning technology \cite{Laviada15}. 

{The numerical simulations are done in 3D, but with a 2D reflector geometry and a SAR trajectory on the plane of the scatterers as illustrated in Figure \ref{fig:geometry}. In this configuration, the horizontal axis shows the range or depth of the scatterers, and the
vertical  axis shows their cross-range.  Because the phases of the signals carry the information from  the scatterers' range, it is usually assumed that range cannot be determined from phaseless data. We see, however, that with the proposed computational imaging method  both range and cross-range are obtained as if full phase information was available.}

We assume that time-resolved intensities can be measured for a set of pulses of the form $\exp(\imath \omega_l t)\exp(-t^2/(2\sigma_t)^2)$, $\omega_l=\omega_1, \omega_2,\dots,\omega_S$, where $\sigma_t$ is the pulsewidth that is inversely proportional to the available bandwidth $B$. As discussed in Appendix \ref{app:illum}, to retrieve the phases for the $S$ frequencies 
we need to measure the intensities for $3S-2$ illuminations. A simple strategy consists of using a sufficient delay between successive illuminations so that the corresponding echoes are non-overlapping. 
Given an illumination at time $0$, an estimate for the starting time for the scattered signal is $\frac{2L}{c}$, while its duration is of the order $\frac{2\,  {\rm IW}_{\rm size} }{c}$, {where ${\rm IW}_{\rm size}$ is the size of the IW}. Thus, an estimate of the acquisition time per location is $(3S-2)\left(\frac{1}{2B}+ \frac{2\,  {\rm IW}_{\rm size}}{c} + \frac{2L}{c}\right)$. For the specific parameters used in the simulations this is about $(3S-2) 10 {\rm ns}=1.2\mu{\rm s}$. 

\begin{figure}[htbp]
\begin{center}
\includegraphics[scale=0.3]{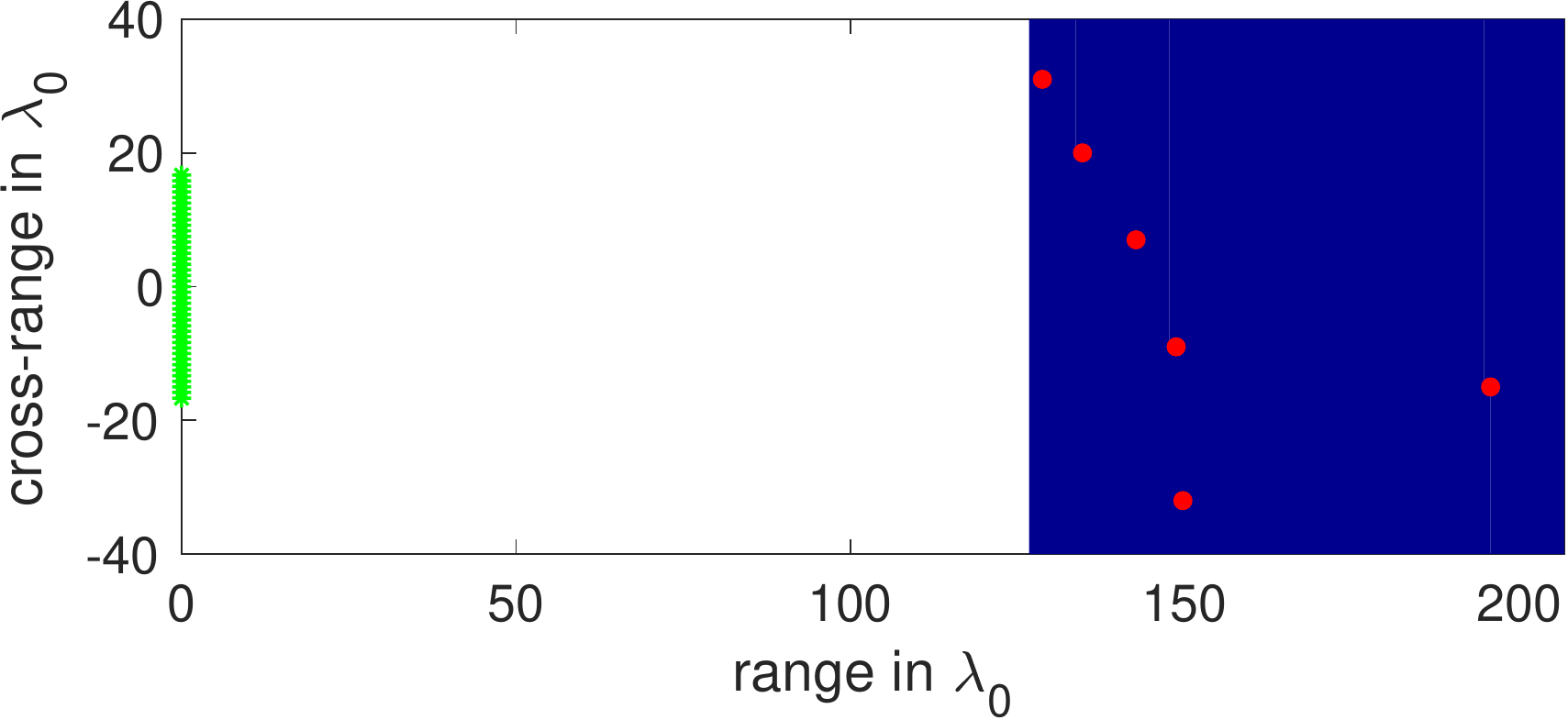} 
\end{center}
\caption{The setup used in the numerical simulations. A single transmit/receive element is moving on a linear trajectory (green stars) and measures the intensity reflected from the scatterers (red disks). The blue area is the imaging window IW.}
\label{fig:geometry}
\end{figure}

 {In our simulations, we add to the signals that arrive to the receiver, including phases, mean zero gaussian noise corresponding to a SNR~$=10$dB. 
Then, their intensities are computed and these are the data from which we form the images} 

Indeed, following the methodology described in Section \ref{sec:holo}, we recover from these intensities the holographic data \eqref{eq:datah}, which have phases that are coherent over frequency and measurement locations. In our numerical examples, we used $\epsilon=0.001\mu{\rm m}$ in \eqref{Rjk}. Note that $\epsilon$ 
is small with respect to the wavelength and the pixel size, so that neglecting distances that differ less than $\epsilon$ does not affect the accuracy of the recovered phases. 
{The Algortihm \ref{algo} is used with 
$\tau=20 \left< \left | \Ac^* \vect b \right| \right>$ and $\beta=\frac{1}{2 \|\Ac \|_2}$.  Here, $\Ac^*$ is the complex conjugate transpose of $\Ac$ and $< \cdot >$ denotes the mean.  
The termination criterium is $\frac{\| \vect z_{k} - \vect z_{k-1} \|_2}{\| \vect z_{k-1} \|_2} \le 1.0e-13$, with $\vect z_{k}$ and $\vect z_{k-1}$ denoting the vector $\vect z$ during the current and the previous iteration.}

The results are shown in Figure \ref{fig:res}. The top row of Figure \ref{fig:res} shows  the distribution of targets we seek to find. The bottom left panel is the KM image, and the right panel is the image obtained with the $\ell_1$-minimization algorithm. As expected, KM shows resolution $\lambda_0 L/a$ in the cross-range direction and $c/B$ in the range direction, which for our imaging setup corresponds to a resolution of $5\lambda_0$ in both directions. On the other hand, the image obtained with the $\ell_1$-minimization algorithm is almost exact.

\begin{figure}[th]
\begin{center}
\includegraphics[scale=0.22]{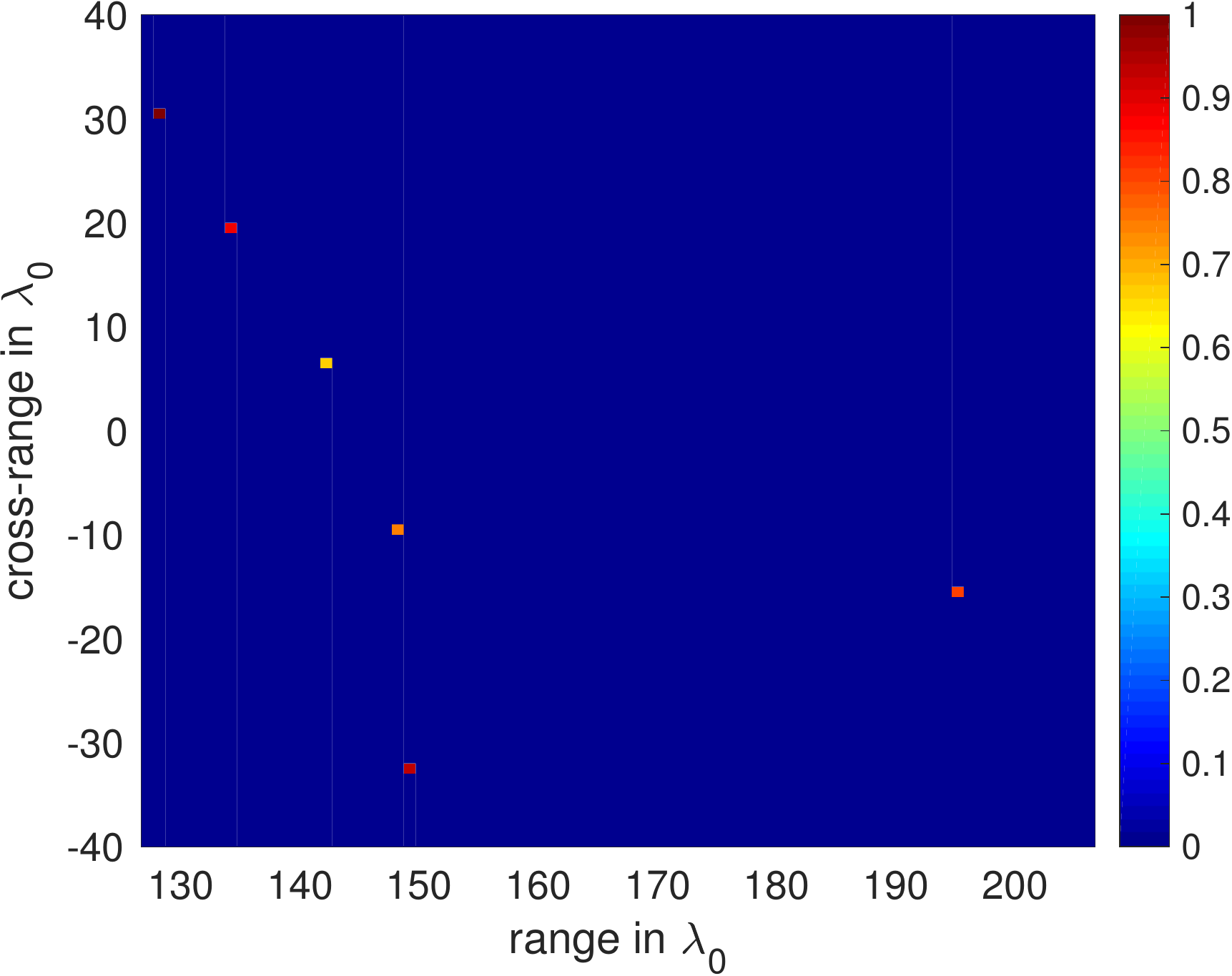} \\
\begin{tabular}{cc}
\includegraphics[scale=0.22]{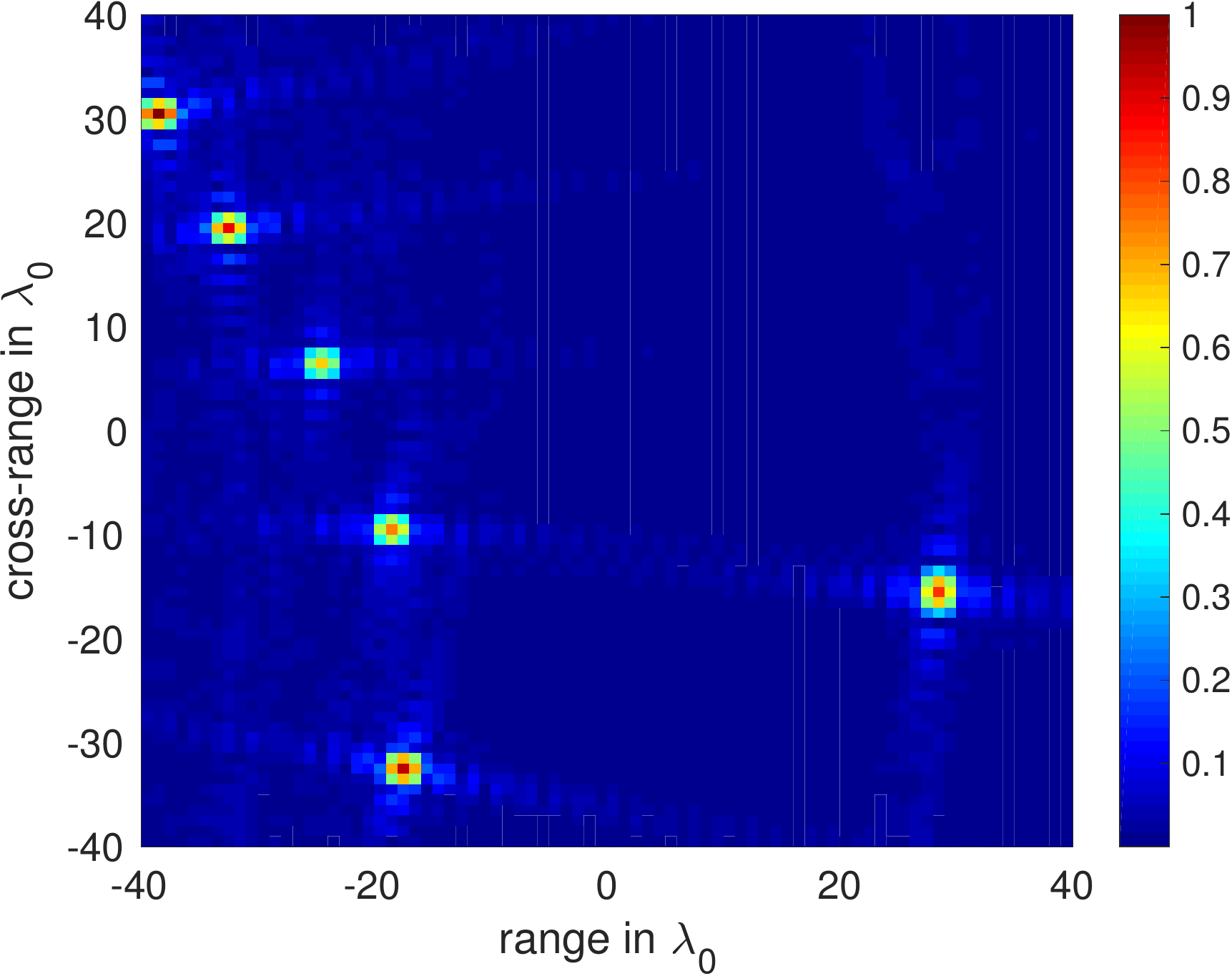}&
\includegraphics[scale=0.22]{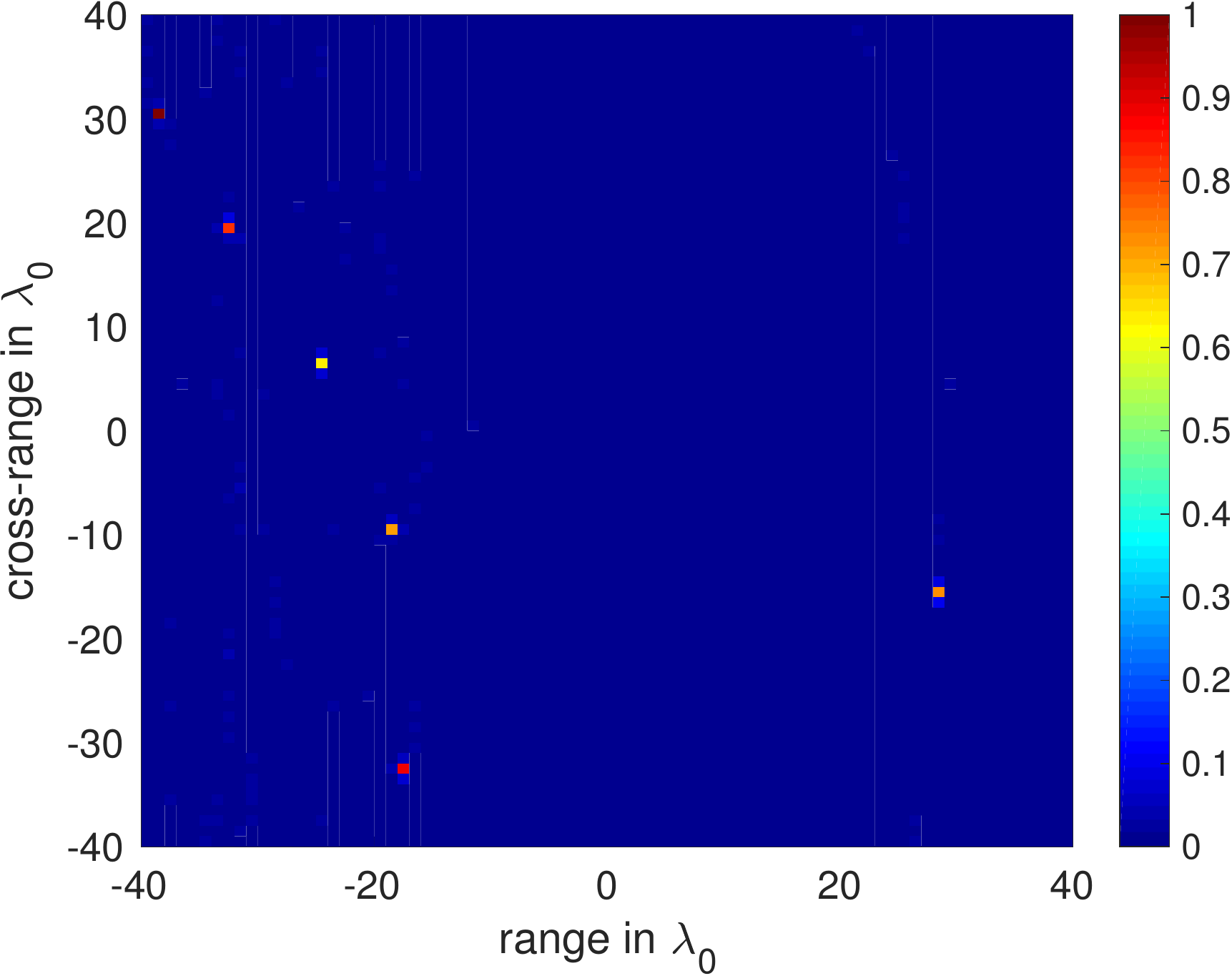} \\
\end{tabular}
\end{center}
\caption{Single transmitter/receiver multifrequency data recovered from intensity measurements with SNR~$=10$dB. Imaging with KM as defined in \eqref{eq:KM_sar} (left) and $\bfrho^{\ell_1}$ computed using GeLMA to solve \eqref{eq:ell1} (right). On the top row the true reflectivity is plotted. In all images we plot the amplitude of the complex valued reflectivity $| \bfrho |$.} 
\label{fig:res}
\end{figure}

 To show the robustness of the  proposed method  when phases are not recorded, we consider next the same imaging configuration but assuming that the scatterers are displaced with respect to the grid points of the IW. Note that misplacements with respect to the grid
amounts to a systematic modelling error that affects the accuracy of the recovered phases and, thus,  deteriorates the  image reconstruction. This is due to errors in the computed distances to the grid points ${R}_k^j$ (see \eqref{dist}) used  in the definition of the radii $r^j_i$ 
in the model matrices \eqref{Acj} and \eqref{Ac}.

  In Figure \ref{fig:res2}, we show results for scatterers that are displaced by $\lambda/8=0.75$mm (top row) and by  $\lambda/2=3$mm (bottom row) with respect to the grid points in both range and cross-range directions. As expected, the reconstructions 
 deteriorate as the displacement with respect to the grid increases, but they remain quite accurate even for the largest possible displacement value of half the grid size; see the bottom row plots in Figure \ref{fig:res2}. 
 
 \begin{figure}[th]
\begin{center}
\begin{tabular}{cc}
\includegraphics[scale=0.22]{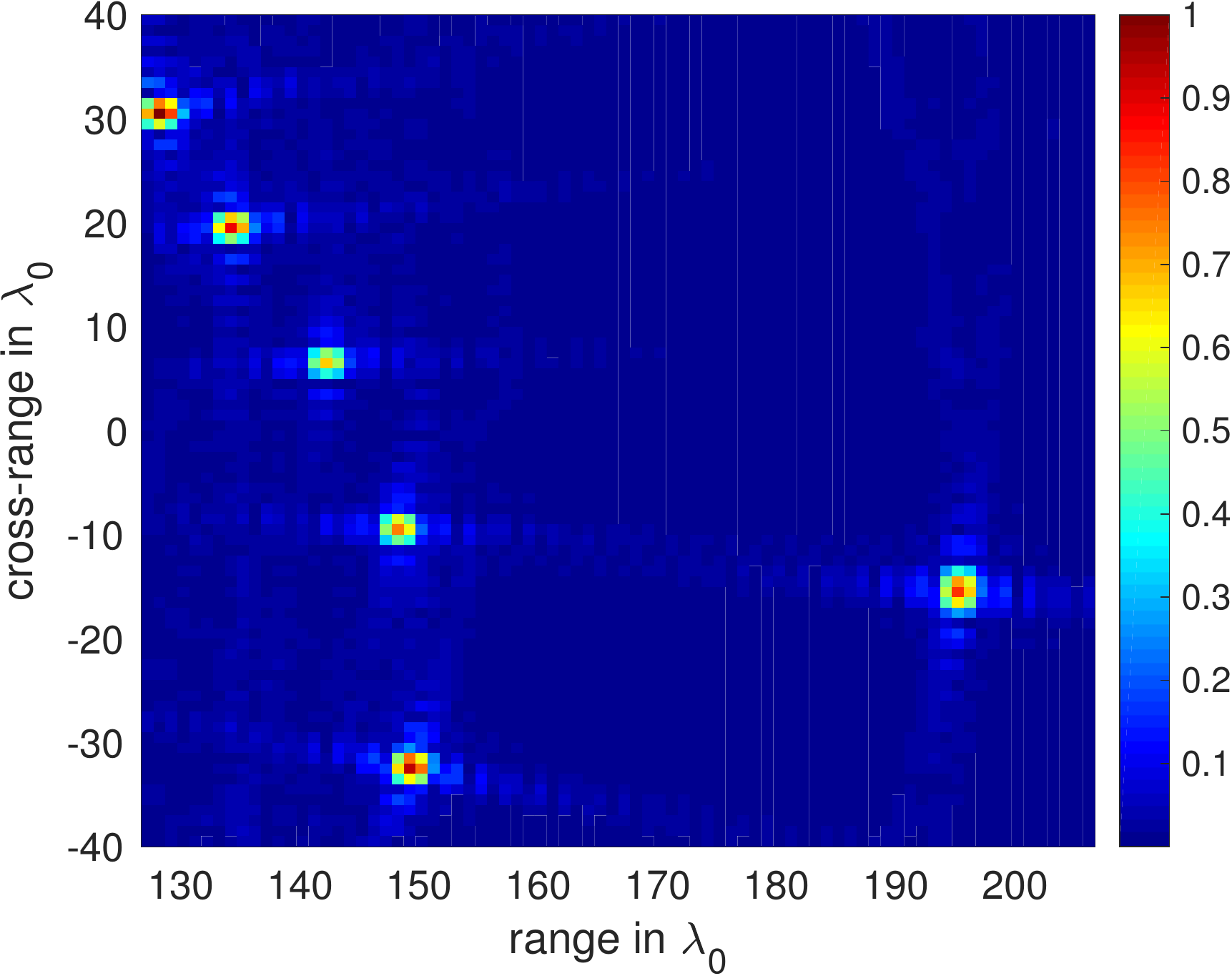}&
\includegraphics[scale=0.22]{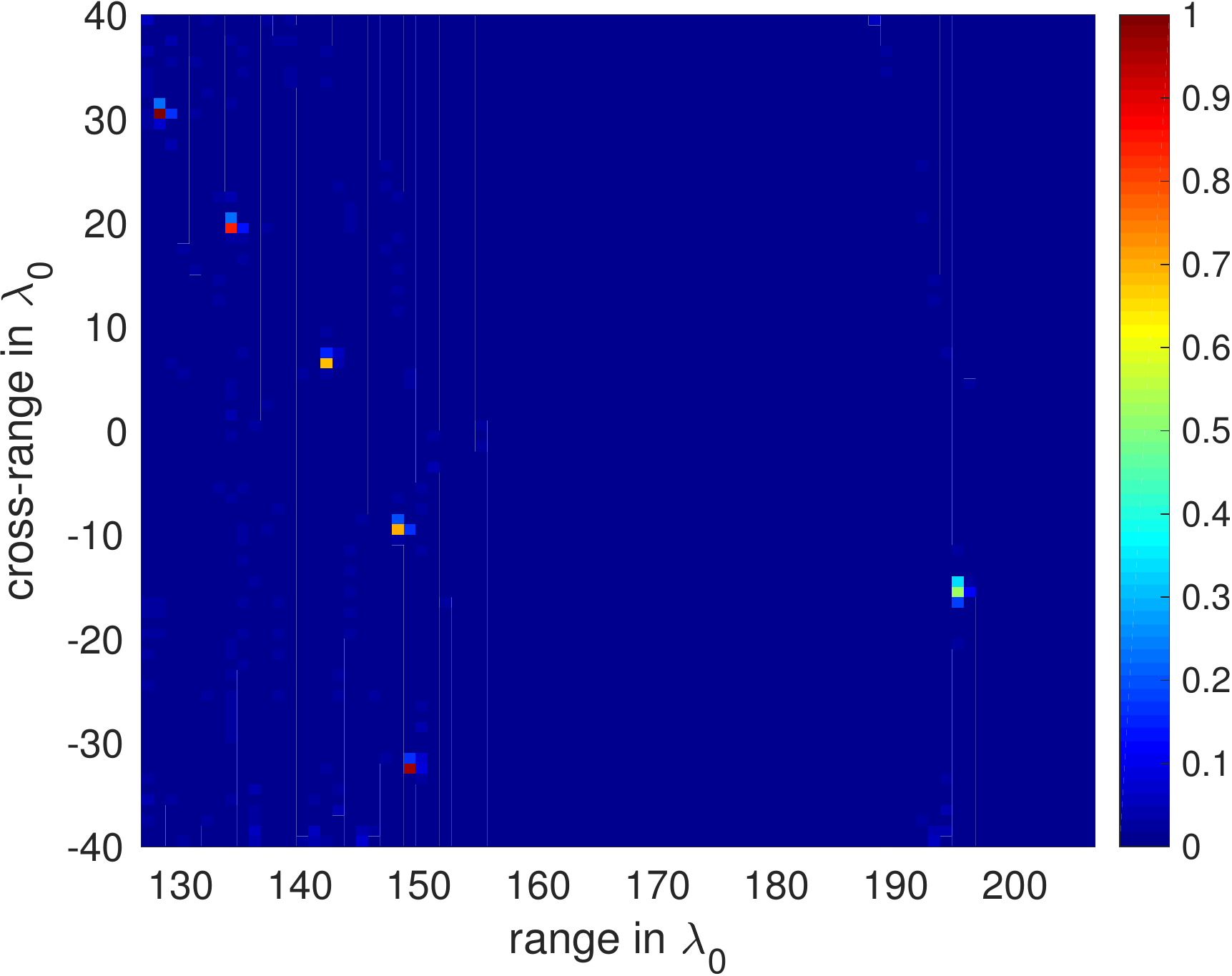} \\
\includegraphics[scale=0.22]{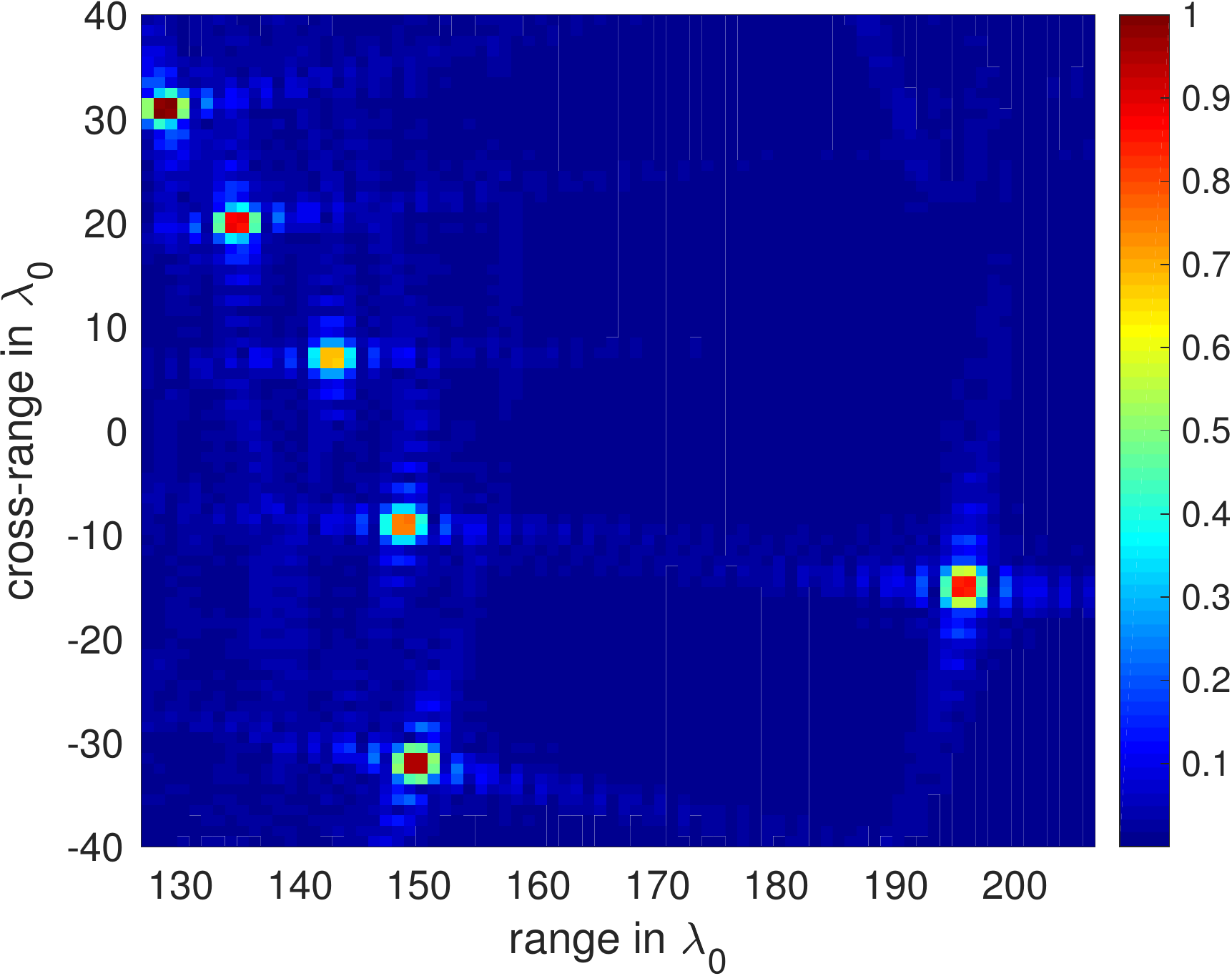}&
\includegraphics[scale=0.22]{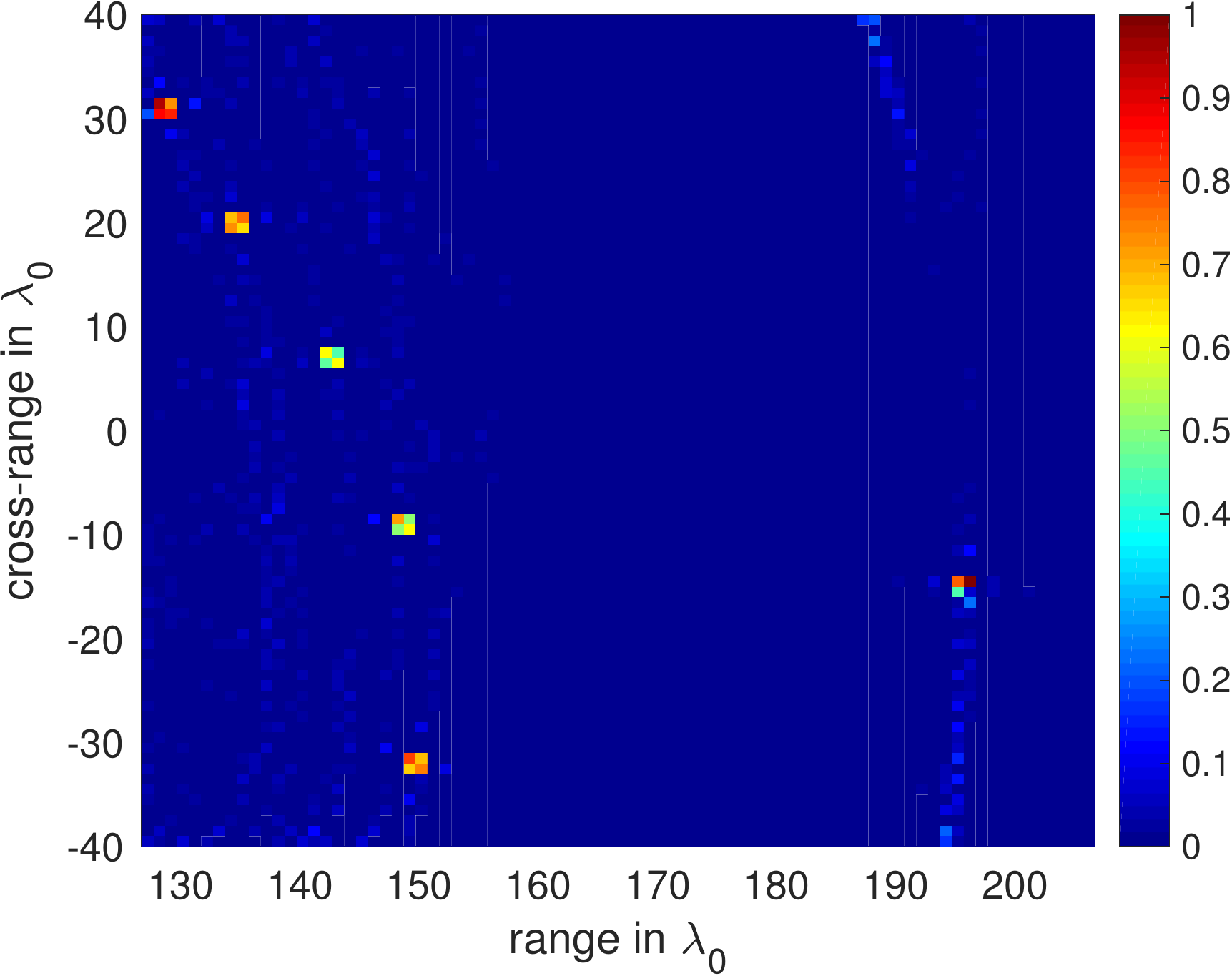} 
\end{tabular}
\end{center}
\caption{Single transmitter/receiver multifrequency data recovered from intensity measurements. No additive noise is added to the data. Imaging with KM as defined in \eqref{eq:KM_sar} (left) and $\bfrho^{\ell_1}$ computed using GeLMA to solve \eqref{eq:ell1} (right). The scatterers are displaced by $\lambda/8=0.75$mm (top row) and by  $\lambda/2=3$mm with respect to the grid points in range and cross-range directions.   In all images we plot the amplitude of the complex valued reflectivity $| \bfrho |$.} 
\label{fig:res2}
\end{figure}

Finally, we also add to the signals that arrive to the receiver corresponding to the displacement $\lambda/2=3$mm, including phases,  mean zero gaussian noise corresponding to a SNR~$=0$dB. The results shown in Figure \ref{fig:res3} are also very good. They illustrate 
the robustness of the proposed methodology with respect to both  additive noise and off-grid displacement errors.

\begin{figure}[th]
\begin{center}
\begin{tabular}{cc}
\includegraphics[scale=0.22]{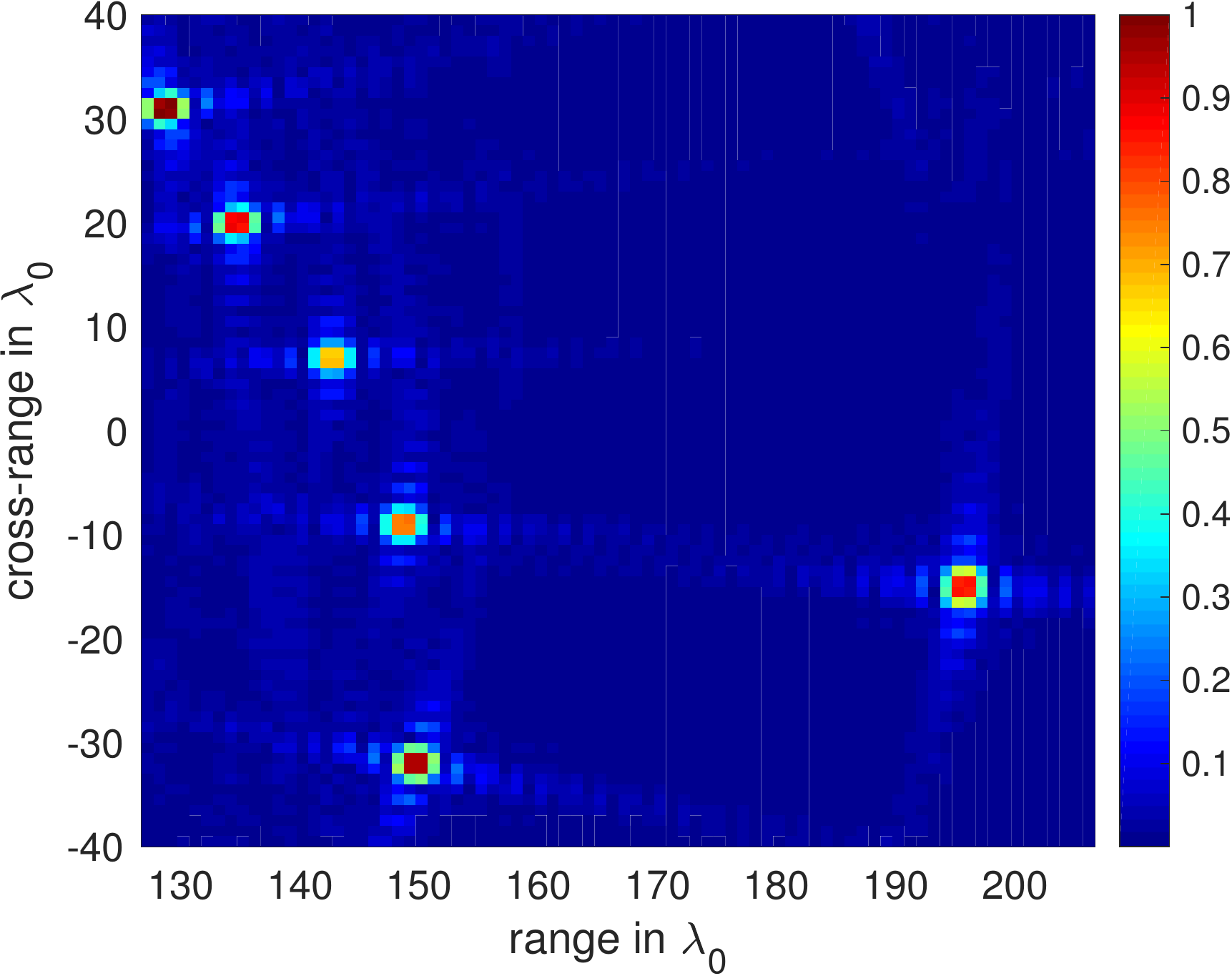}&
\includegraphics[scale=0.22]{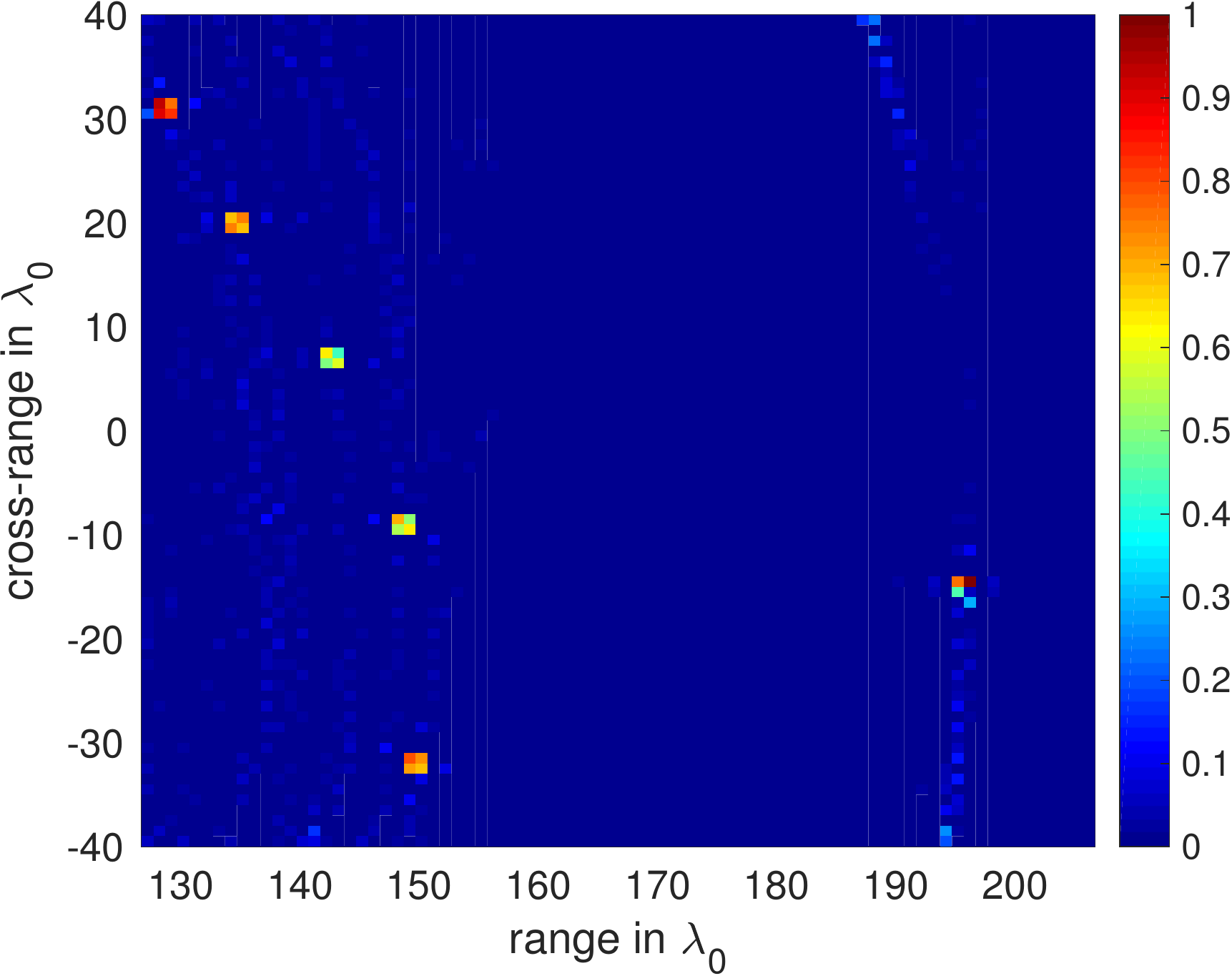} 
\end{tabular}
\end{center}
\caption{Single transmitter/receiver multifrequency data recovered from intensity measurements with SNR~$=0$dB. Imaging with KM as defined in \eqref{eq:KM_sar} (left) and $\bfrho^{\ell_1}$ computed using GeLMA to solve \eqref{eq:ell1} (right). The scatterers are displaced by $\lambda/2=3$mm with respect to the grid points in range and cross-range directions.  In all images we plot the amplitude of the complex valued reflectivity $| \bfrho |$.} 
\label{fig:res3}
\end{figure}

\section{Summary and conclusions}
\label{sec:conclusions}
We have introduced an approach to synthetic aperture imaging with intensity-only measurements that exploits the available diversity in the illuminations. 
The images have the same quality as when full phase information is available.
There are two stages in our approach.  First, we recover {field} cross-correlations over pairs of 
frequencies at each measurement location $\vect x_j$ using intensity-only data. 
Thus, the phases are recovered up to a location dependent factor $e^{\imath \theta_j}$, which is independent of the frequency. 
In the second stage, which is the main contribution of this paper, we introduce an algorithm 
that recovers the missing  phases up to a single global factor $e^{\imath \theta_1}$ for all locations and all frequencies.  
We can then image with any method that uses full (phase and amplitude) data.
We explore this approach with broadband SAR in the $50$GHz regime in an imaging setup that is used in security scanning equipments.

\appendix
\section{Illumination strategies} \label{app:illum}
We discuss here an illumination strategy for recovering the phase cross-correlations from time-resolved intensity measurements at different frequencies using a single transmit/receiver element. 
In this protocol at each transmit/receiver location we need to record measurements corresponding to $3S-2$ illuminations. We explain next the proposed protocol. 

We want to recover the cross-correlation matrix 
\begin{equation}
\label{CC1-app}
  [\Mm^j]_{ll'}\equiv\dsp m^j_{ll'} = \overline{P(\vect x_j,\omega_l)}
  P(\vect x_j,\omega_{l'}),
  \quad l,l'=1,\dots,S\,\, ,
\end{equation}
using diversity of illuminations and the polarization identity
\begin{equation}\label{polarization_re2}
\mbox{Re}(m^j_{ll'})= 
\frac{1}{2} \left( \norm{\bP^j \cdot \we_{l+l'}}^2  - 
\norm{\bP^j \cdot \we_l}^2 - \norm{\bP^j \cdot \we_{l'}}^2\right)
\end{equation}
\begin{equation}\label{polarization_im2}
\mbox{Im}(m^j_{ll'})= 
 \frac{1}{2} \left( \norm{\bP^j \cdot \we_{l-{\imath} l'}}^2  -
  \norm{\bP^j \cdot  \we_l}^2 - \norm{\bP^j \we_{l'}}^2\right). 
  \end{equation}
When  intensities are recorded, all the quantities $\norm{ \cdot }^2$ in the right hand side of \eqref{polarization_re2} and \eqref{polarization_im2} are known. 

Remarking that 
$$ m^j_{ll'} = \frac{m^j_{l1} m^j_{1l'}}{m^j_{11}}\,, $$
we deduce that we only need to compute the phase cross-correlations $m^j_{l1} m^j_{1l'}$ which can be obtained from the polarization identity \eqref{polarization_re2}-\eqref{polarization_im2} provided $3 S-2$ measurements. Indeed we can determine $m^j_{l1}$, for $l=2,\ldots,S$ using illuminations $\we_{l}$, $\we_{l+1}$ and   $\we_{l-{\imath}1}$ and we also need to measure $m^j_{11}$. 

{
\section{Measurement process} \label{app:meas}
We describe here the measurement process that allows us to compute the elements $m^j_{ll'} $ of the cross-correlation matrix $\Mm^j$ corresponding to frequencies $\omega_l$ and $\omega_{l'}$. 
When two signals of distinct frequencies $\omega_l$ and $\omega_{l'}$ are sent simultaneously to probe the medium, the intensity at the receivers oscillates at the difference frequency 
$\omega_l - \omega_{l'}$. Therefore, our imaging system should be able to resolve intensities that oscillate at frequencies of the order of the total bandwidth $B$.
}
{
Indeed, the intensity of the slowly varying terms is of the form 
\begin{eqnarray*}\label{oscillation}
I_0 &=& \left| P(\vect x_j,\omega_l) \right|^2 + \left| P(\vect x_j,\omega_{l'}) \right|^2 \\ 
&+& 2 \left| P(\vect x_j,\omega_l) \right| \left| P(\vect x_j,\omega_{l'}) \right| \cos[(\omega_l-\omega_{l'})t +\Delta\phi^j_{ll'}]\, , \\ 
I_{\pi/2} &=& \left| P(\vect x_j,\omega_l) \right|^2 + \left| P(\vect x_j,\omega_{l'}) \right|^2 \\ 
&+& 2 \left| P(\vect x_j,\omega_l) \right| \left| P(\vect x_j,\omega_{l'}) \right| \sin[(\omega_l-\omega_{l'})t +\Delta\phi^j_{ll'}]\, .
\end{eqnarray*}
We remark that both $I_0$ and $I_{\pi/2}$ oscillate with a slow frequency $\omega_l - \omega_{l'}$. We assume that we can measure the integral of these terms over some time interval $\Delta t \ll \frac{1}{B}$.
This leads to a linear system of the form 
\begin{equation}
\label{phase_diff} 
\left[
\begin{array}{ll} 
 \alpha_1 & -\alpha_2 \\
 \alpha_2 & \alpha_1 
\end{array}
 \right] 
  \left[ 
 \begin{array}{l} 
 \cos(\Delta\phi^j_{ll'}) \\ 
 \sin(\Delta\phi^j_{ll'}) 
 \end{array}
 \right] = 
 \left[
  \begin{array}{l} 
 \beta_1 \\
 \beta_2
\end{array}
\right]
\end{equation}
with 
$$\alpha_1=\int_{t_0}^{t_0+\Delta t} \cos(  (\omega_l-\omega_{l'})t ) dt,\ \alpha_2=\int_{t_0}^{t_0+\Delta t} \sin(  (\omega_l-\omega_{l'})t ) dt$$
$$\beta_1=\int_{t_0}^{t_0+\Delta t} \frac{I_0 - \left| P(\vect x_j,\omega_l) \right|^2 - \left| P(\vect x_j,\omega_{l'}) \right|^2 }{2 \left| P(\vect x_j,\omega_l) \right| \left| P(\vect x_j,\omega_{l'}) \right| } dt$$
$$\beta_2=\int_{t_0}^{t_0+\Delta t} \frac{I_{\pi/2} - \left| P(\vect x_j,\omega_l) \right|^2 - \left| P(\vect x_j,\omega_{l'}) \right|^2 }{2 \left| P(\vect x_j,\omega_l) \right| \left| P(\vect x_j,\omega_{l'}) \right| } dt.$$
By solving system (\ref{phase_diff}) subject to the constraint 
$$ \cos^2(\Delta\phi^j_{ll'})+\sin^2(\Delta\phi^j_{ll'})=1$$
we determine the phases  $\Delta\phi^j_{ll'}$ of the elements $m^j_{ll'}= |m^j_{ll'}| e^{\Delta\phi^j_{ll'}}$, $l\neq l'$. 
Then, it is straightforward to obtain the phase differences
}

\section*{Acknowledgments}
The work of M. Moscoso was partially supported by Spanish grant MICINN FIS2016-77892-R. The work of A.Novikov was partially supported by NSF grant NSF DMS-1813943. The work of G. Papanicolaou was partially supported by AFOSR FA9550-18-1-0519. The work of C. Tsogka was partially supported by AFOSR FA9550-17-1-0238 and AFOSR FA9550-18-1-0519.

%

\end{document}